\documentclass[preprint,preprintnumbers,nobibnotes,pra,showkeys,onecolumn,secnumarabic,showpacs]{revtex4}
\usepackage{amssymb}
\usepackage{amsmath}
\usepackage{amsfonts}
\usepackage{graphicx}
\usepackage{dcolumn}
\usepackage{bm}

\input{tcilatex}
\begin{document}

\title{Quantum quench dynamics in XY spin chain with ferromagnetic and
antiferromagnetic interactions }
\author{Zhe Wang,$^{1,2}$ Pan-Pan Fang,$^{2}$ Yu-Liang Xu,$^{1}$ Chun-Yang
Wang,$^{1}$ Rong-Tao Zhang,$^{2}$ Han Zhang$,^{2}$ }
\author{Xiang-Mu Kong$^{1,2}$}
\altaffiliation{Corresponding author. E-mail address: kongxm668@163.com (X.-M. Kong).}
\affiliation{$^{1}$School of Physics and Optoelectronic Engineering, Ludong University,
Yantai 264025, China\\
$^{2}$College of physics and Engineering, Qufu Normal University, Qufu
273165, China }

\begin{abstract}
In this manuscript we investigate the one-dimensional anisotropic XY model
with ferromagnetic and antiferromagnetic interactions, which gives more
interesting phase diagrams and dynamic critical behaviors. By using quantum
renormalization-group method, we find that there are three phases in the
system: antiferromagnetic Ising phase ordered in \textquotedblleft $x$
direction\textquotedblright , spin-fluid phase and ferromagnetic Ising phase
ordered in \textquotedblleft $y$ direction\textquotedblright . In order to
study the dynamical critical behaviors of the system, two quantum quenching
methods are used. In both cases, the concurrence, a measure of entanglement,
oscillates periodically over time. We show that the periods are the same and
can be used as a new order parameter for quantum phase transitions. For
further discussion, we derive the scaling exponent, $\theta $, and
correlation length exponent, $\nu $, from the scaling behavior of the
evolution period.
\end{abstract}

\maketitle

\section{INTRODUCTION}

As a kind of nonlocal correlation, quantum entanglement has attracted
extensive attention in many study fields. In quantum information theory, it
can be regarded as an essential resource for processing information that is
impossible to achieve in classical information theory\cite{1,2,3,4}. In
condensed matter physics, it\ plays a key role in the study of quantum phase
transition(QPT) which occurs at zero temperature and is induced by the
changing of parameters in the Hamiltonian \cite{5,6,7}.

In the past few years, the static properties between entanglement and the
QPT in quantum many-body spin systems are studied widely. For example, the
Ising, XY, and XXZ models are investigated by using quantum renormalization
group(QRG) method. In these studies it was shown that the entanglement can
be used as an order parameter to QPT and the critical exponents can be
derived from its scaling behavior\cite{8,9,10,11,12,13,14}.

Recently as the advance in ultracold atomic, molecule experiments, and the
development of ultrafast pulsed lasers\cite{15,16,17,18}, there is a growing
savor for studying the dynamics of quantum many-body spin systems\cite%
{19,20,21,22,23}. Among them, Jafari\ study the dynamics of the
one-dimensional Ising model in transverse field by QRG method, and find that
the characteristic time(at which the entanglement reaches its 1th maximum)
can used to define the critical point\cite{19}.\ Hazzard \textit{et al}.
investigate XXZ spin system following a quantum quench which is easily to
achieve experimentally and reveal that time evolution of correlation
function manifests a nonperturbative dynamic singularity\cite{21}. Meng Qin 
\textit{et al}. discuss the non-equilibrium evolution of quantum coherence
in one-dimensional anisotropic XY model by giving an initial state
arbitrarily and find that quantum coherence periodically fluctuates over time%
\cite{23}.

However, there are some open questions still need to answer. Can we study
the dynamic critical behaviors of the quantum many-body spin systems deeply
by quantum quench? Can we find a new order parameter for quantum phase
transition? If there are new order parameters, what critical information can
we derive from it? Based on these questions, we study the dynamics of
quantum entanglement for one-dimensional ferromagnetic-antiferromagnetic
spin-$1/2$ XY system. The organization of this manuscript is as follows.\ In
Sec. II the phase diagram of the system is discussed by QRG method. In Sec.
III the dynamic evolution of the system by two quantum quenching methods is
studied. The relation between the evolution period of entanglement and the
QPT is investigated in sec. IV and the summary is given in the last section.
\ 

\section{QRG METHOD AND PHASE DIAGRAM}

The XY model was initially studied by E. Lieb and since then it was widely
investigated\cite{10,11,23,24,25,26,27}. In this manuscript we go on to
study the XY spin chain with ferromagnetic and antiferromagnetic
interactions. The Hamiltonian of the system on a periodic chain with $N$
sites reads%
\begin{equation}
H=\frac{1}{4}\underset{i=1}{\overset{N}{\sum }}\left( J_{x}\sigma
_{i}^{x}\sigma _{i+1}^{x}-J_{y}\sigma _{i}^{y}\sigma _{i+1}^{y}\right) ,
\label{1}
\end{equation}%
where $\sigma _{i}^{\alpha }\left( \alpha =x,y\right) $ are Pauli operators
at site $i$,$\ J_{x}=$ $J\left( 1+\gamma \right) $ and\ $J_{y}=J\left(
1-\gamma \right) \left( J>0\right) $\ are exchange coupling constants, in
which $-1\leq \gamma \leq 1$ is the anisotropy parameter.

QRG method is well suited to solve many-body systems, and they are
conceptually easy to be extended to the higher dimensions. Next, we study
the phase diagram of this system by Kadanoff's block QRG method\cite%
{28,29,30,31}. Its main idea is to reduce the degrees of freedom until a
manageable situation is reached. A brief introduction to the steps of QRG is
follows\cite{8,9,10,11}. First and foremost is that the lattice are divided
into blocks in a reasonable way. Secondly, the block Hamiltonian is
diagonalized exactly. After that, the low-lying eigenstates of all the
blocks are used to construct projection operatorscons. By it the full
Hamiltonian is projected onto these eigenstates, which gives the effective
Hamiltonian. Final, the QRG equations are obtained by calculating and
comparing between previous Hamiltonian and the effective one. The advantage
of the QRG method is not only to simplify the system but also to keep the
low energy spectrum of the system invariant, which is very important to
explore quantum issues with zero temperature.

Above one has briefly introduced the steps of QRG, and will apply it to our
system. One find that three sites as a block is most reasonable, and based
on Kadanoff's block approach we write the Eq. (1) as%
\begin{equation}
H=H^{\text{B}}+H^{\text{BB}},
\end{equation}%
where $H^{\text{B}}$ is block Hamiltonian and $H^{\text{BB}}$\ is
interblock\ Hamiltonian. The specific forms of $H^{\text{B}}$\ and $H^{\text{%
BB}}$ are%
\begin{equation}
H^{\text{B}}=\sum_{I=1}^{N\diagup 3}h_{I}^{\text{B}},
\end{equation}%
in which%
\begin{equation}
h_{I}^{\text{B}}=\left[ J_{x}\left( \sigma _{I,1}^{x}\sigma
_{I,2}^{x}+\sigma _{I,2}^{x}\sigma _{I,3}^{x}\right) -J_{y}\left( \sigma
_{I,1}^{y}\sigma _{I,2}^{y}+\sigma _{I,2}^{y}\sigma _{I,3}^{y}\right) \right]
/4,
\end{equation}%
\begin{equation}
H^{\text{BB}}=\sum_{I=1}^{N\diagup 3}\left( J_{x}\sigma _{I,3}^{x}\sigma
_{I+1,1}-J_{y}\sigma _{I,3}^{y}\sigma _{I+1,1}^{y}\right) /4,
\end{equation}%
where $\sigma _{I,j}^{\alpha }\left( \alpha =x,y\right) $ refer to the Pauli
matrix at site $j$ of the block labeled by $I.$

In the space spanned by $\left\{ \left\vert \uparrow \uparrow \uparrow
\right\rangle ,\left\vert \uparrow \uparrow \downarrow \right\rangle
,\left\vert \uparrow \downarrow \uparrow \right\rangle ,\left\vert \uparrow
\downarrow \downarrow \right\rangle ,\left\vert \downarrow \uparrow \uparrow
\right\rangle ,\left\vert \downarrow \uparrow \downarrow \right\rangle
,\left\vert \downarrow \downarrow \uparrow \right\rangle ,\left\vert
\downarrow \downarrow \downarrow \right\rangle \right\} $\ $\left( \text{%
where }\left\vert \uparrow \right\rangle \ \text{and }\left\vert \downarrow
\right\rangle \text{\ represent the eigenstates of }\sigma ^{z}\right) ,$\
the block Hamiltonian can be exactly diagonalized and the degenerate ground
states can be obtained as%
\begin{equation}
\left\vert \Phi _{0}\right\rangle =\frac{1}{2\sqrt{1+\gamma ^{2}}}\left( -%
\sqrt{1+\gamma ^{2}}\left\vert \uparrow \uparrow \downarrow \right\rangle +%
\sqrt{2}\gamma \left\vert \uparrow \downarrow \uparrow \right\rangle -\sqrt{%
1+\gamma ^{2}}\left\vert \downarrow \uparrow \uparrow \right\rangle +\sqrt{2}%
\left\vert \downarrow \downarrow \downarrow \right\rangle \right) ,
\label{2}
\end{equation}%
\begin{equation}
\left\vert \Phi _{0}^{%
{\acute{}}%
}\right\rangle =\frac{1}{2\sqrt{1+\gamma ^{2}}}\left( -\sqrt{2}\left\vert
\uparrow \uparrow \uparrow \right\rangle -\sqrt{2}\gamma \left\vert
\downarrow \uparrow \downarrow \right\rangle +\sqrt{1+\gamma ^{2}}\left\vert
\uparrow \downarrow \downarrow \right\rangle +\sqrt{1+\gamma ^{2}}\left\vert
\downarrow \downarrow \uparrow \right\rangle \right) .  \label{3}
\end{equation}%
The degenerate ground states of all blocks are used to build projection
operator $T_{0}$ where $T_{0}=\tprod\nolimits_{I=1}^{N/3}T_{0}^{I},$ and
specific form of $T_{0}^{I}$\ is%
\begin{equation}
T_{0}^{I}=\left\vert \Uparrow \right\rangle _{I}\left\langle \Phi
_{0}\right\vert +\left\vert \Downarrow \right\rangle _{I}\left\langle \Phi 
{\acute{}}%
_{0}\right\vert ,
\end{equation}%
in which $\left\vert \Uparrow \right\rangle _{I}$\ and $\left\vert
\Downarrow \right\rangle _{I}$\ are renamed base kets in the effective
subspace. Projection operators can be used to associate effective
Hamiltonian with initial one in the following form 
\begin{equation}
H^{\text{eff}}=T^{\dag }H^{\text{B}}T+T^{\dag }H^{\text{BB}}T.
\end{equation}%
By calculating and comparing with Eq. (1), we get the effective one as%
\begin{equation}
H^{\text{eff}}=\underset{I=1}{\overset{N/3}{\sum }}\left[ J_{x}^{^{\prime
}}\sigma _{I}^{x}\sigma _{I+1}^{x}-J_{y}^{\prime }\sigma _{I}^{y}\sigma
_{I+1}^{y}\right] /4  \label{4}
\end{equation}%
with%
\begin{equation}
J_{x}^{^{\prime }}=J^{\prime }\left( 1+\gamma ^{^{\prime }}\right)
,J_{y}^{\prime }=J^{\prime }\left( 1-\gamma ^{^{\prime }}\right) ,
\end{equation}%
in which

\begin{equation}
J^{^{\prime }}=J\frac{3\gamma ^{2}+1}{2\left( 1+\gamma ^{2}\right) },\gamma
^{^{\prime }}=\frac{\gamma ^{3}+3\gamma }{3\gamma ^{2}+1}.
\end{equation}%
\begin{figure}[tbp]
\centering
\includegraphics[width =160 mm, height = 80 mm]{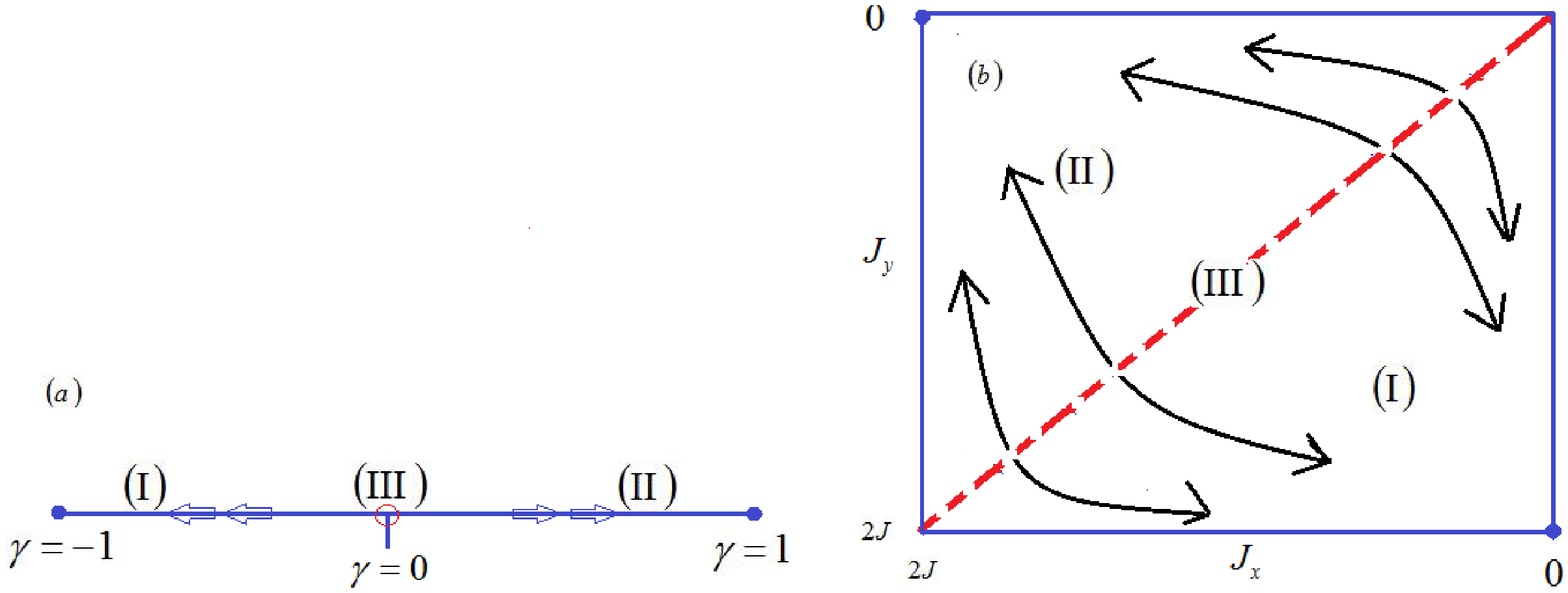}\newline
\caption{$\left( \text{Coloronline}\right) $\ Phase diagram of the
one-dimensional ferromagnetic-antiferromagnetic spin-1/2 XY system. Arrows
show running of couplings under QRG. Filled circles represent stable fixed
points and open circle is unstable fixed point and the red dotted line
corresponds to the open circle.$\ $Phase$\left( \text{I}\right) $\ is
ferromagnetic\ Ising phase ordered in \textquotedblleft y
direction\textquotedblright , $\left( \text{II}\right) $\ represents
antiferromagnetic Ising phase ordered in \textquotedblleft x
direction\textquotedblright , and $\left( \text{III}\right) $\ is spin-fluid
phase.}
\label{3}
\end{figure}
Above equation is the ultimate results of QRG and from it we can get a lot
of information of the system. Based on it, we plot the the phase diagrams of
the system in Fig. 1 and we will give a simple explanation following. At
zero temperature, a phase transition occurs upon relative variation of the
parameters in the Hamiltonian\cite{32}. In this system the exchange coupling
constant $J_{x}$ makes the spins staggered arrangement along the
\textquotedblleft x direction\textquotedblright\ i.e., makes the system in
antiferromagnetic Ising phase. The exchange coupling constant $-J_{y}$ makes
all spins in the \textquotedblleft positive or negative direction of
y\textquotedblright\ i.e., makes the system in ferromagnetic Ising phase.
The competition between $J_{x}$ and $J_{y}$ leads to the QPT between
antiferromagnetic Ising phase and ferromagnetic one. An advantage of this
model is that the competition can be also described by a single $\gamma $.
From Eq. (1) we can see that when $\gamma $ determined $J_{x}$ and $J_{y}$
can be determined simultaneously. So the QPT can be described by $J_{x}$ and 
$J_{y}$\ or $\gamma .$

For simplicity, let's analyze the phase diagram of the system with a single
gama, and correspond to a two-dimensional parameter space. By solving $%
\gamma ^{^{\prime }}=\gamma $, we get the fixed points in which the stable
one are $\gamma =\pm 1$ and the unstable one is $\gamma =0.$\ At $\gamma =1,$%
\ the system is in the antiferromagnetic Ising phase ordered in
\textquotedblleft x direction\textquotedblright .\ The $\gamma =-1$
corresponds to ferromagnetic Ising phase ordered in \textquotedblleft $y$
direction\textquotedblright . At $\gamma =0,$ the spins can point to either
\textquotedblleft x directions\textquotedblright\ or\ \textquotedblleft y
directions\textquotedblright\ due to the competition between $J_{x}$ and $%
J_{y}$ is the same. For this point,the system is in the spin-fluid phase,
because the most disordered state but largest entanglement which will show
in in the Sec.III\cite{33}. Other points flow to corresponding stable fixed
point at the thermodynamic limit, which tells us that the $\gamma _{c}=0$\
is the transition point from ferromagnetic Ising phase $\left( 0>\gamma
\geqslant -1\right) $ to the spin-fluid phase $\left( \gamma =0\right) $\ or
from spin-fluid phase to the antiferromagnetic Ising phase $\left( 1\geq
\gamma >0\right) .$

\section{QUENCH DYNAMICS \label{A ZA}}

There are many interesting and unknown things in dynamic critical properties
of quantum many-body spin systems. Exploring these things is important to
understand the nature of quantum systems. In this section, we study the
dynamic critical behaviors of entanglement in
ferromagnetic-antiferromagnetic spin-1/2 XY system by two quantum quenching
methods(1. closing the external field suddenly; 2. changing the coupling
constant suddenly) and we use concurrence to measure entanglement \cite%
{34,35,36,37}.

Quench 1: From the experimental point of view, one can make the system with
Hamiltonian 
\begin{equation*}
H_{1}=\sum_{i=1}^{N}\left( J_{x}\sigma _{i}^{x}\sigma _{i+1}^{x}-J_{y}\sigma
_{i}^{y}\sigma _{i+1}^{y}+h\sigma _{i}^{z}\right) /4
\end{equation*}
in the ground state by low temperature. When the system is in the ground
state, we can realize quench by removing transverse field, $h$, abruptly at
time $t=0$. In short, the system with Hamiltonian $H$ begins to evolve from
ground state of the Hamiltonian $H_{1}$. All in all, this method is easy to
implement.

In the process of dynamics, it is difficult to calculate the concurrence of
large system directly. Next we first calculate the concurrence of the
three-site system and then investigate it in a large one by QRG method. The
calculation process is as follows. The density matrix of the three-site
system before quenching needs to be given first by $\rho _{1}\left( 0\right)
=\left\vert \varphi _{_{1}}\right\rangle \left\langle \varphi
_{_{1}}\right\vert $ in which $\left\vert \varphi _{_{1}}\right\rangle $ =$%
\left\vert \downarrow \downarrow \downarrow \right\rangle $\.{ }Then the
density matrix at time $t$ can be gotten by $\rho \left( t\right) =U\left(
t\right) \rho \left( 0\right) U^{\dag }\left( t\right) $ where $U\left(
t\right) =e^{-iHt}$ is unitary operator. In the space which has been given
in Sec. II, the unitary operator is obtained as%
\begin{equation}
U\left( t\right) =\left( 
\begin{array}{cccccccc}
a & 0 & 0 & b & 0 & c & b & 0 \\ 
0 & d & e & 0 & f & 0 & 0 & b \\ 
0 & e & g & 0 & e & 0 & 0 & c \\ 
b & 0 & 0 & d & 0 & b & f & 0 \\ 
0 & f & e & 0 & d & 0 & 0 & b \\ 
c & 0 & 0 & b & 0 & g & e & 0 \\ 
b & 0 & 0 & f & 0 & e & d & 0 \\ 
0 & b & c & 0 & b & 0 & 0 & a%
\end{array}%
\right) ,
\end{equation}%
in which we consider $J$=1 and the specific values are as follows%
\begin{eqnarray*}
a &=&\frac{\gamma ^{2}+\cos \omega t}{1+\gamma ^{2}},b=-\frac{i\sin \omega t%
}{\sqrt{2\left( 1+\gamma ^{2}\right) }}, \\
c &=&\frac{\gamma \left( -1+\cos \omega t\right) }{1+\gamma ^{2}},d=\left(
\cos \frac{\omega t}{2}\right) ^{2}, \\
e &=&-\frac{i\gamma \sin \omega t}{\sqrt{2\left( 1+\gamma ^{2}\right) }},f=%
\frac{1}{2}\left( -1+\cos \omega t\right) , \\
g &=&\frac{1+\gamma ^{2}\cos \omega t}{1+\gamma ^{2}},\omega =\frac{\sqrt{%
1+\gamma ^{2}}}{\sqrt{2}}.
\end{eqnarray*}%
After get the density matrix at time t, we can get pairwise concurrence by
tracing the degrees of freedom of one site. Without loss generality, we
trace site 2 and the reduced density matrix for sites 1 and 3 at time $t$\
can be obtained as%
\begin{equation}
\rho _{1}\left( \gamma ,t\right) =\left( 
\begin{array}{cccc}
k & 0 & 0 & q \\ 
0 & m & m & 0 \\ 
0 & m^{\ast } & m & 0 \\ 
q^{\ast } & 0 & 0 & p%
\end{array}%
\right) ,  \label{9}
\end{equation}%
where%
\begin{eqnarray*}
k &=&\frac{\left( \gamma ^{2}+\cos \omega t\right) ^{2}}{\left( 1+\gamma
^{2}\right) ^{2}}, \\
q &=&\frac{\gamma \left( -1+\cos \omega t\right) \left( \gamma ^{2}+\cos
\omega t\right) }{\left( 1+\gamma ^{2}\right) ^{2}}, \\
m &=&\frac{\left( \sin \omega t\right) ^{2}}{2+2\gamma ^{2}}, \\
p &=&\frac{\gamma ^{2}\left( -1+\cos \omega t\right) ^{2}}{\left( 1+\gamma
^{2}\right) ^{2}}.
\end{eqnarray*}%
The\ reduced\ density matrix has X-state formation due to $k+2m+p=1,$ $%
m^{\ast }=m,$ and $q^{\ast }=q$. Based on this, we get the concurrence as%
\cite{38}\ 
\begin{equation}
C_{1}\left( \gamma ,t\right) =2\max \left\{ 0,m-\sqrt{kp},\left\vert
q\right\vert -m\right\} .  \label{10}
\end{equation}

Quench 2: The system is initially in the ground state of the Hamiltonian $%
H_{2}=\sum_{i=1}^{N}\left[ J_{x}\sigma _{i}^{x}\sigma _{i+1}^{x}+J_{y}\sigma
_{i}^{y}\sigma _{i+1}^{y}\right] /4$ and then the $J_{y}$ is changed
suddenly to $-J_{y}$ at $t=0$.\ The initial state is $\left\vert \varphi
_{2}\right\rangle =\frac{1}{2}\left( -\left\vert \uparrow \uparrow
\downarrow \right\rangle +\frac{\sqrt{2}}{\sqrt{1+\gamma ^{2}}}\left\vert
\uparrow \downarrow \uparrow \right\rangle -\frac{1}{\sqrt{2}}\left\vert
\downarrow \uparrow \uparrow \right\rangle +\frac{\sqrt{2}\gamma }{\sqrt{%
1+\gamma ^{2}}}\left\vert \downarrow \downarrow \downarrow \right\rangle
\right) .$ Similarly, the reduced density matrix for sites 1 and 3 can be
gotten as%
\begin{equation}
\rho _{2}\left( \gamma ,t\right) =\left( 
\begin{array}{cccc}
x & 0 & 0 & s \\ 
0 & z & z & 0 \\ 
0 & z^{\ast } & z & 0 \\ 
s^{\ast } & 0 & 0 & y%
\end{array}%
\right) ,  \label{11}
\end{equation}%
where%
\begin{eqnarray*}
s &=&-\frac{\gamma -10\gamma ^{3}+\gamma ^{5}-4\gamma \left( -1+\gamma
^{2}\right) \cos \omega t+\gamma \left( -1+\gamma ^{2}\right) ^{2}\cos
2\omega t-2i\left( -1+\gamma ^{2}\right) \left( 1+\gamma ^{2}\right)
^{2}\sin \omega t}{4\left( 1+\gamma ^{2}\right) ^{3}}, \\
x &\text{=}&\frac{2-3\gamma ^{2}+8\gamma ^{4}+\gamma ^{6}+\gamma ^{2}\left(
-1+\gamma ^{2}\right) \left( -8\cos \omega t-\left( -1+\gamma ^{2}\right)
\cos 2\omega t\right) }{4\left( 1+\gamma ^{2}\right) ^{3}}, \\
y &=&\frac{1+8\gamma ^{2}-3\gamma ^{4}+2\gamma ^{6}+8\gamma ^{2}\left(
-1+\gamma ^{2}\right) \cos \omega t-\left( -1+\gamma ^{2}\right) ^{2}\cos
2\omega t}{4\left( 1+\gamma ^{2}\right) ^{3}}, \\
z &=&\frac{1+6\gamma ^{2}+\gamma ^{4}+\left( -1+\gamma ^{2}\right) \cos
2\omega t}{8\left( 1+\gamma ^{2}\right) ^{2}},
\end{eqnarray*}%
and the concurrence can be obtained as%
\begin{equation}
C_{2}\left( \gamma ,t\right) =\text{2Max}\left\{ 0,z-\sqrt{xy},\left\vert
s\right\vert -z\right\} .  \label{12}
\end{equation}%
\begin{figure}[tbp]
\centering
\includegraphics[width =160 mm, height = 60 mm]{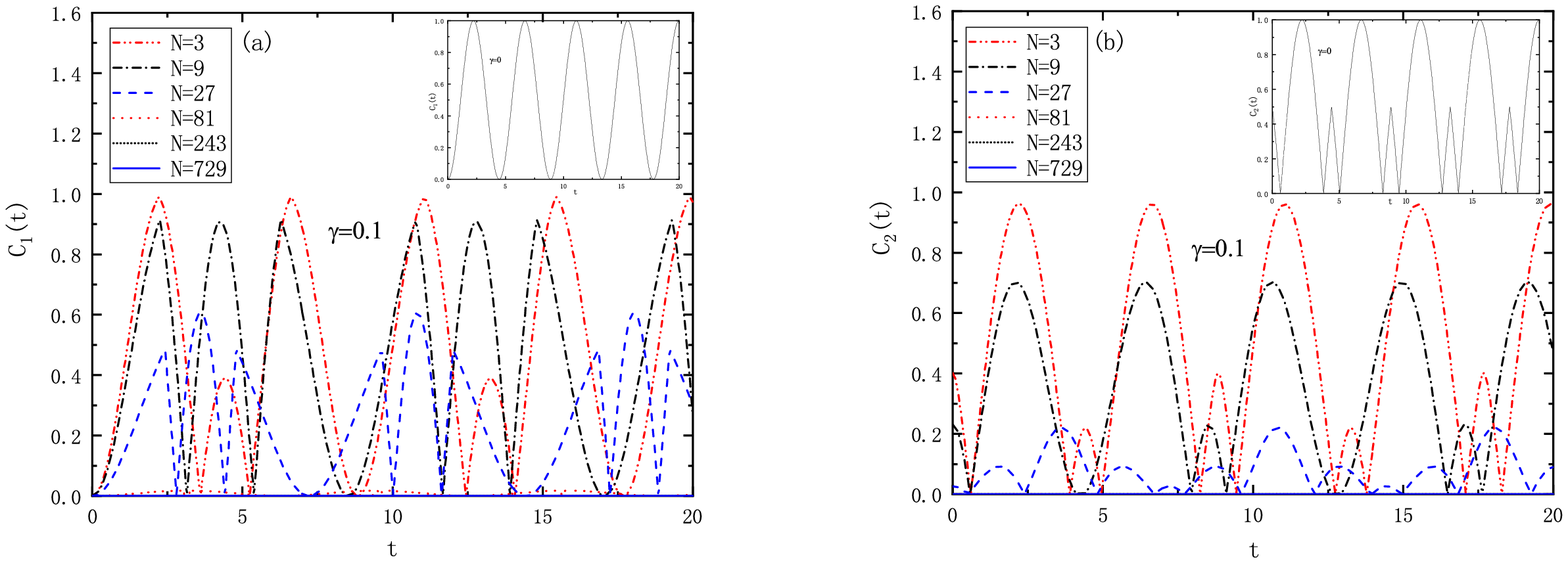}\newline
\caption{$\left( \text{Coloronline}\right) $ Evolution of concurrences over
the time under different QRG steps at $\protect\gamma =0.1$. $\left( \text{a}%
\right) $\ and $\left( \text{b}\right) $\ correspond to the first and second
kind of quantum quench respectively. Inset: Concurrences of different size
of the system coincide with each other at the critical$\ \left( \protect%
\gamma =0\right) .$ }
\label{3}
\end{figure}
\begin{figure}[tbp]
\centering
\includegraphics[width =160 mm, height = 60 mm]{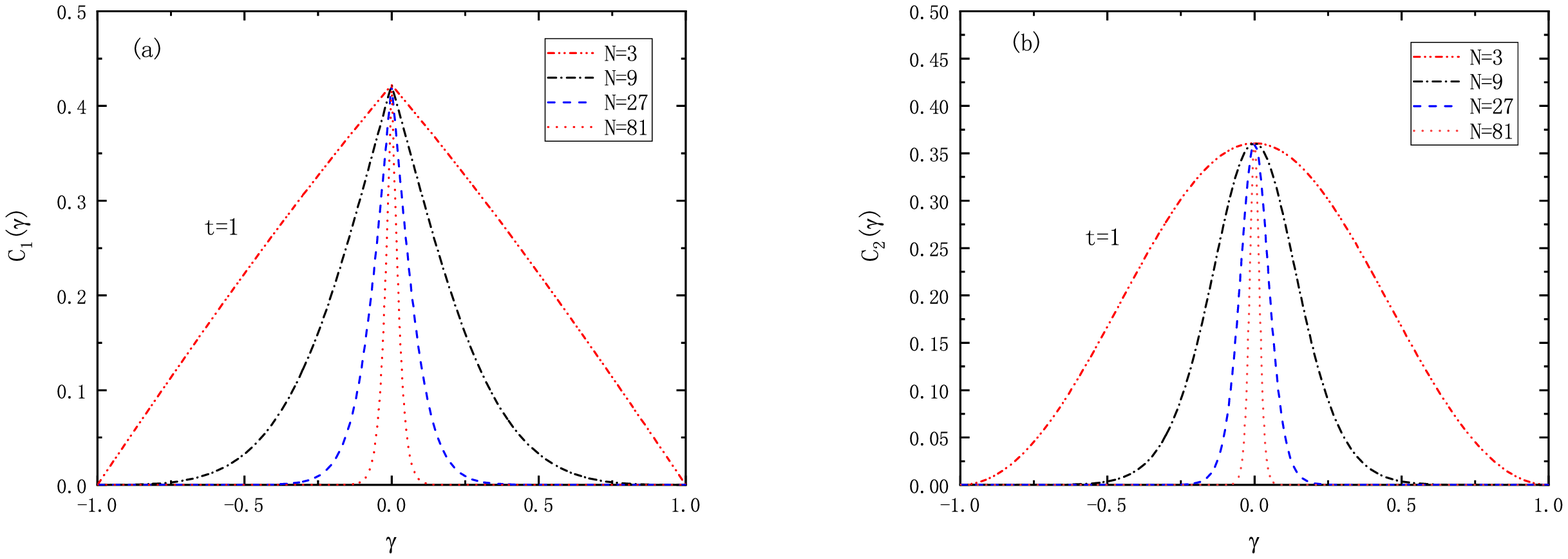}\newline
\caption{$\left( \text{Coloronline}\right) $ Variation of the concurrences\
versus $\protect\gamma $ after both kinds of quantum quench at $t=1.$\ As
the size of the system increases, the concurrences show singular behavior at 
$\protect\gamma =0$ which shows that the QPT occurs here.}
\label{3}
\end{figure}
For more intuitive analysis, we first show the evolution of concurrences
over the time under QRG steps in the Fig. 2. (a) describes the first quench
method, (b) describes the second one. From Fig. 2$\left( \text{a}\right) ,$\
we find that the concurrence periodically oscillates from zero. This may be
caused by the constant fluctuation between the ground state of the system
before quenching and after quenching. As the size of the system increases,
the concurrence decreases gradually and disappears in a large system. The
reason for this is that any points except the critical point flow to a
stable fixed point $\left( \text{after\ enough\ iterations}\right) $ and the
state corresponding to it can be separated into the product of subsystems.
Moreover, the concurrence of different-length chains coincide with each
other at the critical point $\left( \text{Fig. 2, inset}\right) $, because
that the correlation length diverges at this point. From Fig. 2$\left( \text{%
b}\right) $, we can see that the concurrence periodically oscillates from
no-zero values, which indicates that the initial state is entangled.
Comparing the two figures,we can see that the performance of concurrence are
different when system is finite,but the characteristic is consistent under
the thermodynamic limit.In other words,different quenching methods do not
affect the critical phenomenon,which can also be shown by the scaling
behavior of the same period later. 

\begin{figure}[tbp]
\centering
\includegraphics[width =160 mm, height =100 mm]{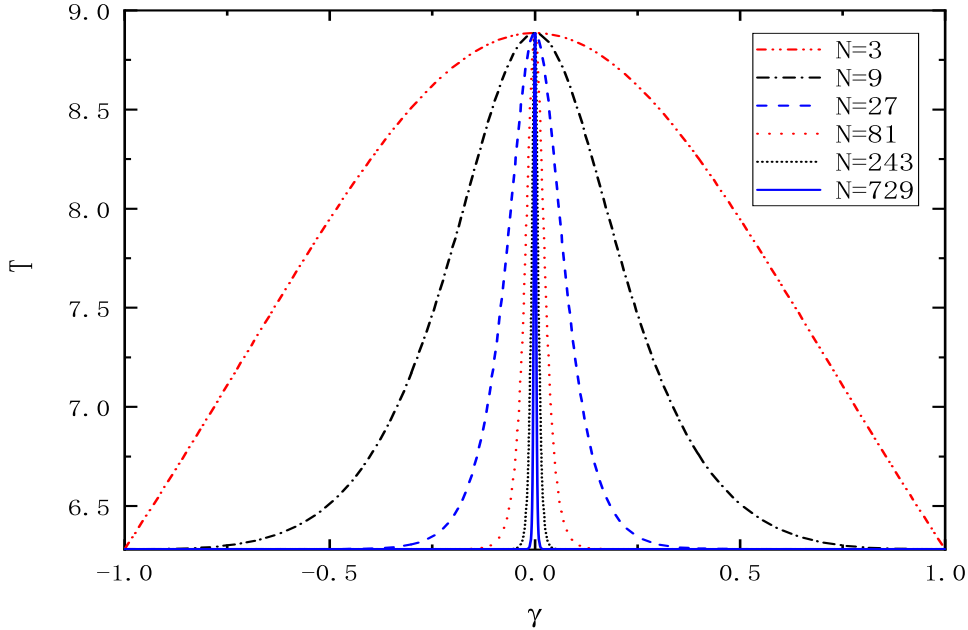}\newline
\caption{ $\left( \text{Coloronline}\right) $ Evolution period $T$ as a
function of $\protect\gamma $ for different QRG steps. Similar to
concurrence, the period $T$ shows singular behavior at $\protect\gamma =0.$}
\label{3}
\end{figure}
\begin{figure}[tbp]
\centering
\includegraphics[width =160 mm, height =100 mm]{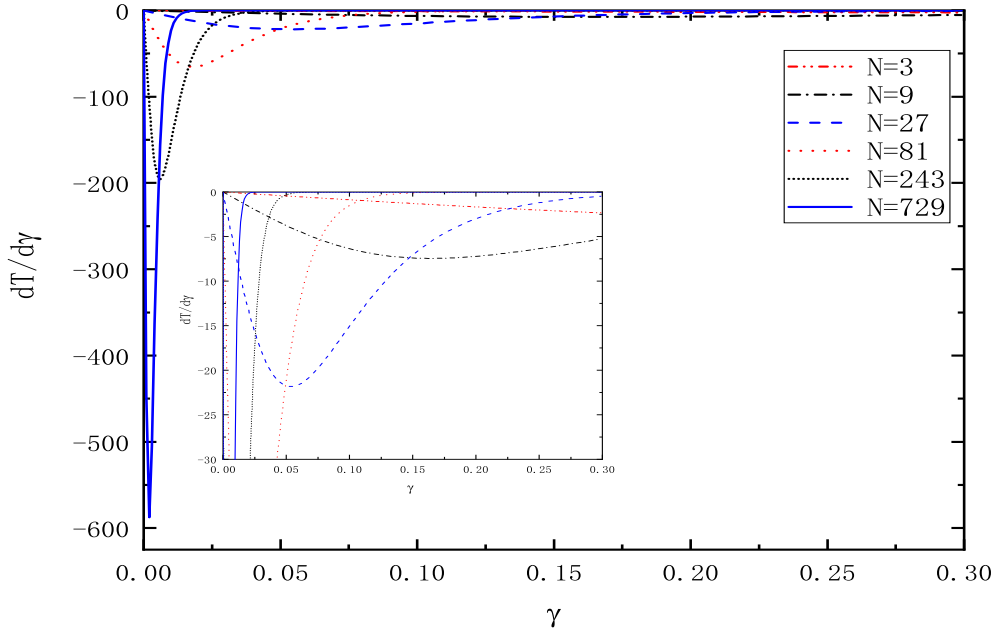}\newline
\caption{ $\left( \text{Coloronline}\right) $ First derivation of the
evolution period $T$\ with respect to the $\protect\gamma $ for different
size of the system. In the figure, we only give the $\protect\gamma \geq 0$,
because the $T$ is an even function of $\protect\gamma .$}
\label{3}
\end{figure}
\begin{figure}[tbp]
\centering
\includegraphics[width =160 mm, height =100 mm]{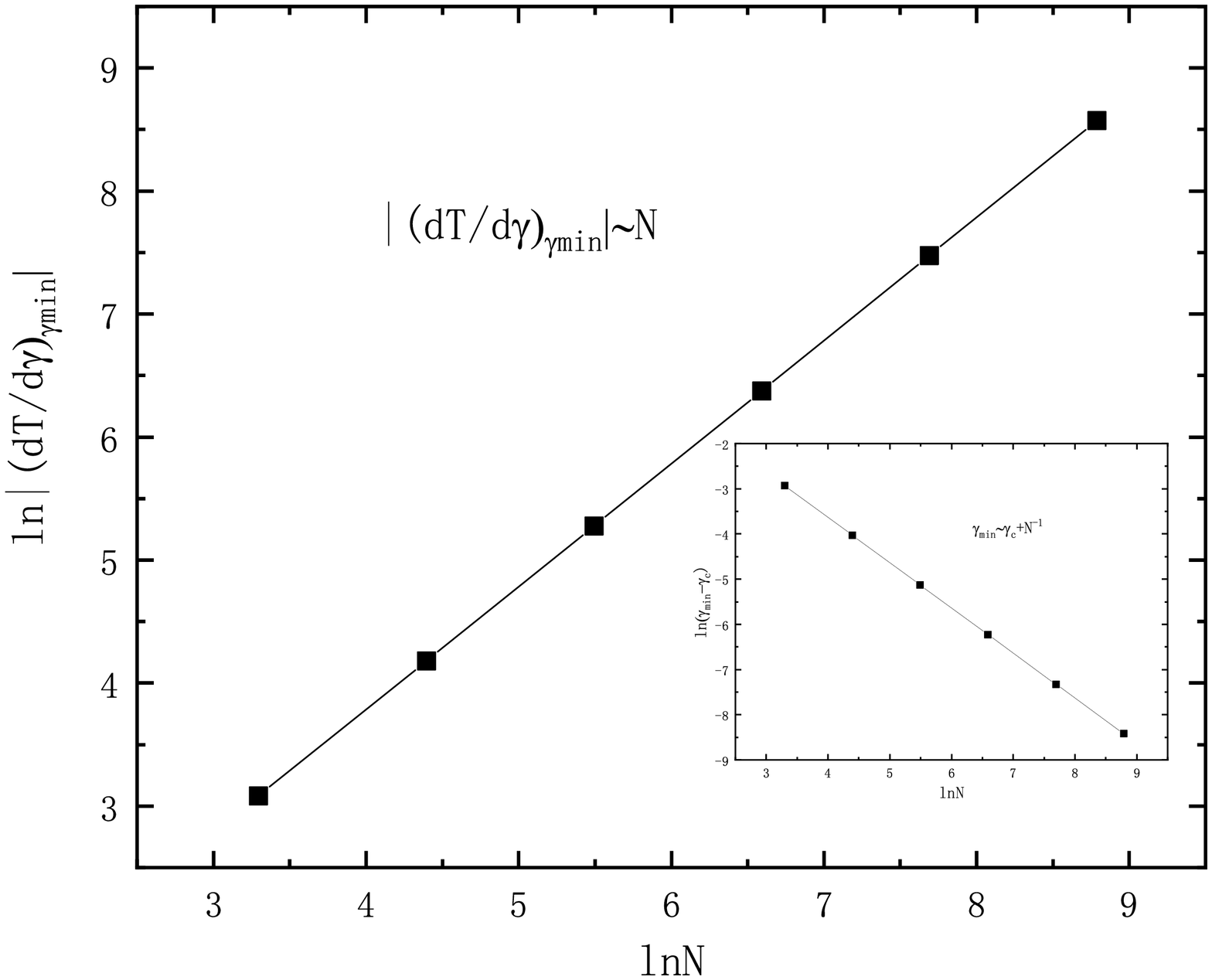}\newline
\caption{ Scaling behavior of the system, ln$\left\vert \left( dT/d\protect%
\gamma \right) _{\min }\right\vert $ with the logarithm of the system size $%
N.$ Inset: Scaling of the position $\left( \protect\gamma _{m}\right) $ of $%
\left( dT/d\protect\gamma \right) _{\min }$ for different-length chains.}
\label{3}
\end{figure}

When time is given the concurrences\ versus $\gamma $ are plotted in Fig. 3.
For the both kinds of quantum quench, it is found that the plots under
different QRG steps cross each other at the critical point which is due to
that this point is fixed point. In the thermodynamic limit, the concurrences
develop two saturated values, one is non-zero corresponding to the ordered
phase and one is zero corresponding to the disordered phase.

\section{EVOLUTION PERIOD AND QPT\label{A ZA copy(1)}}

Above we find that the concurrences periodically oscillate over the time for
both kinds of quantum quench and we deduce that the periods is the same from
Eq. $\left( 15\right) $ and Eq. $\left( 17\right) .$The evolution period, $T,
$ of three-site system is given as%
\begin{equation}
T=2\sqrt{2}\pi /\left( \sqrt{1+\gamma ^{2}}\right) ,
\end{equation}%
and we will study it in a large system with renormalized coupling constants. 

We expect that the evolution period is a new order parameter, because it
does not depend on the quench methods. $T$ with $\gamma $ for different
system sizes, $N,$\ is plotted in Fig. 4. Similar to concurrence, the period 
$T$\ tends to two different values in the thermodynamic corresponding to the
ordered phase and disordered phase respectively. 

For further discussion, we plot the first derivative of the evolution
period, $dT/d\gamma ,$\ in Fig. 5. The derivative of the $T$ diverges at $%
\gamma _{c}=0$ under the thermodynamic limit, which shows\ that a
second-order QPT occurs at this point, which can also be proven by the
scaling behavior of $\left\vert \left( dT/d\gamma \right) _{\min
}\right\vert $ . We have derived the scaling behavior of $\left\vert \left(
dT/d\gamma \right) _{\min }\right\vert $ versus \textit{N }and displayed it
in Fig. 6 which shows a linear behavior of ln$\left\vert \left( dT/d\gamma
\right) _{\min }\right\vert $ versus ln\textit{N.}\ The scaling behavior is $%
\left\vert \left( dT/d\gamma \right) _{\min }\right\vert \sim N^{\theta },$
with exponent $\theta =1$. The exponent $\theta $\ is related to the
correlation length exponent\ closed to the critical point and we can give a
simple derivation process here.\ In the vicinity of $\gamma _{c}=0$, the
correlation length exponent, $\nu ,$ gives the behavior of correlation
length, that is $\xi \sim \left\vert \gamma -\gamma _{c}\right\vert ^{-\nu }.
$ Under the $n$th QRG steps, the correlation length can be written as $\xi
_{\left( n\right) }=\xi /3^{n}\sim \left\vert \gamma _{\left( n\right)
}-\gamma _{c}\right\vert ^{-\nu }$\ where $\gamma _{\left( n\right) }$\ is
renormalized anisotropy parameter and dividing it to $\xi \sim \left\vert
\gamma -\gamma _{c}\right\vert ^{-\nu }$\ gives $\left\vert \left( d\gamma
_{n}/d\gamma \right) \right\vert _{\gamma _{c}}\sim N^{1/\nu }.$\ We can
give $\left\vert \left( dT/d\gamma \right) \right\vert _{\min }\sim
\left\vert \left( d\gamma _{n}/d\gamma \right) \right\vert _{\gamma _{c}}$
since the position of the minimum $\left( \gamma _{\min }\right) $ tends
toward the $\gamma _{c}$ as $\gamma _{\min }=\gamma _{c}+N^{-1}$\ (Fig. 6,
inset). To summarize the above derivation, we can get $\left\vert \left(
dT/d\gamma \right) \right\vert _{\min }\sim N^{\theta }\sim \left\vert
\left( d\gamma _{n}/d\gamma \right) \right\vert _{\gamma _{c}}\sim N^{1/\nu }
$ which implies that $\nu =1/\theta $\cite{8}. The correlation length can be
given as $\xi \sim \left( \gamma -\gamma _{c}\right) ^{-1},$ which shows
that the correlation length diverges at $\gamma _{c}=0.$

\section{\protect\bigskip SUMMARY \label{A ZB}}

The phase diagram and the dynamic critical behaviors of the concurrence in
one-dimensional ferromagnetic-antiferromagnetic spin-1/2 XY system are
studied in this manuscript. By the competition analysis between $J_{x}$ and $%
J_{y},$ the phases at the fixed points are obtained. The phase diagram of
the system is given base on renormalization equation. We discuss the
dynamics of the concurrences by both kinds of quantum quench which is easy
to achieve experimentally and we find that the concurrences periodically
oscillate over the time. The periods are the same and can be seen as order
parameter to QPT. Finally, we get the correlation length exponent by
analyzing the scaling behavior of the evolution period. The correlation
length exponent equals to one, which indicates that the correlation length
diverges at $\gamma _{c}=0$.

\bigskip\ \ \ 

\begin{center}
\bigskip ACKNOWLEDGMENTS
\end{center}

This work is supported by the National Natural Science Foundation of China
under Grants NO. 11675090, NO. 11505103, and NO. 11847086. Z.W. would like
to thank Zhong-Qiang Liu, Yong-Yang Liu, Xiao-Hui Jiang, Qi-Ming Wang,
Li-Yuan Wang, and Pan-Pan Zhang for fruitful discussions and useful comments.

\begin{center}
\bigskip
\end{center}

\end{document}